\documentclass{appolb}
\usepackage{epsfig}

\usepackage{pslatex}

\usepackage[cp1250]{inputenc}
\usepackage[T1]{fontenc}
\usepackage{amsmath}
\newcommand{\bs}[1]{\,\boldsymbol{#1}}

\begin{document}
\title{Synchronization of Coupled Oscillators \\ in a Local One-Dimensional Kuramoto Model
\thanks{Presented at Summer Solstice 2009 International Conference on Discrete Models of Complex Systems}
}
\author{
J. Ochab
\address{Marian Smoluchowski Institute of Physics, Jagellonian University\\ Reymonta 4, 30-059 Kraków, Poland}
\and
P.F. Góra
\address{Marian Smoluchowski Institute of Physics \\and Mark Kac Complex Systems Research Centre,\\Jagellonian University, Reymonta 4, 30-059 Kraków, Poland}
}
\maketitle
\begin{abstract}
A modified Kuramoto model of synchronization in a finite discrete system of locally coupled oscillators is studied. The model consists of N oscillators with random natural frequencies arranged on a ring. It is shown analytically and numerically that finite-size systems may have many different synchronized stable solutions which are characterised by different values of the winding number. The lower bound for the critical coupling $k_c$ is given, as well as an algorithm for its exact calculation. It is shown that in general phase-locking does not lead to phase coherence in 1D.
\end{abstract}
  
\section{Introduction}
The phenomenon of collective synchronization, in which a large population of elements oscillating with different frequencies spontaneously locks to a common frequency, is the subject of extensive research in physics, biology, chemistry, and social sciences. Biological examples at various levels of complexity contain pacemaker cells in the heart \cite{serce1,serce2}, the metabolic synchrony in yeast cell suspensions \cite{drozdze1,drozdze2} or the synchronously flashing swarm of fireflies \cite{swietliki1,swietliki2}. The studies of synchronizing man-made systems include arrays of lasers \cite{lasery1,lasery2} or superconducting Josephson junctions \cite{josephson1,josephson2}.

The Kuramoto model \cite{Kuramoto}, which treats the system as an ensemble of limit-cycle oscillators described only by their phases, proved to be a very successful approach to the problem of synchronization. The natural frequencies of the elements of the population are drawn from some prescribed distribution. It has been shown analytically that the system with a mean-field coupling exhibits a phase transition in the thermodynamic limit: if the coupling exceeds a certain threshold, a macroscopic set of the oscillators spontaneously synchronize.

On the other hand, as shown by Strogatz and Mirollo \cite{Strogatz1}, an infinite 1D system does not synchronize. However, in real-life application only finite systems appear. It is, therefore, interesting to see whether finite 1D Kuramoto systems may synchronize.

\section{A Local One-Dimensional Kuramoto Model}
The topology of interactions of the studied system forms a ring, i.e. we assume the lattice to be one-dimensional with nearest-neighbour coupling, and periodic boundary conditions. The local Kuramoto model of synchronizing limit-cycle oscillators is given by the equations of motion (first-order, nonlinear ODEs)
\begin{equation}
\label{eq:motion1}
\dot{\theta_i}(t)=\omega_i - \frac{k}{2}\left[\sin(\theta_i(t)-\theta_{i-1}(t))+\sin(\theta_i(t)-\theta_{i+1}(t))\right] \,,
\end{equation}
for $i=1,\,\ldots,\,N$, where $\theta_i$ are the phases of the oscillators, $\omega_i$ are the natural frequencies taken from a certain distribution, and $k$ is the coupling constant.

\section{The form of stable synchronized solutions}
We look for stationary synchronized solutions given by
\begin{equation}
\theta_i(t)=\Omega t+\phi_i \quad, i=1,\,\ldots,\,N \quad,
\end{equation}
where $\Omega$ is the common frequency, and the set of $N$ constant phases $\{\phi_i\}_{i=1,\ldots,N}$ characterises a given solution. The insertion of this condition into the equations of motion (\ref{eq:motion1}) yields a linear system of equations for sines of constant phase differences:
\begin{equation}
\label{eq:matrixeq}
\bs{As}=\frac{2}{k}(\bs{\omega}-\Omega \bs{e})\equiv \frac{2}{k}\bs{\Delta} \quad,
\end{equation}
where $\bs{e}=[1,1,\ldots,1]^T$, $\bs{s}=[\sin(\phi_1-\phi_N),\sin(\phi_2-\phi_1),\ldots,\sin(\phi_N-\phi_{N-1})]$, $\bs{\omega}=[\omega_1,\omega_2,\ldots,\omega_N]^T$, $ \bs{\Delta}=[\delta_1,\delta_2,\ldots,\delta_N]^T\equiv[\omega_1-\Omega,\omega_2-\Omega,\ldots,\omega_N-\Omega]^T$ are the deviations of natural frequencies from the mean frequency, and the matrix $\bs{A}$ is given by
\begin{equation}
\bs{A}=
\left[ \begin{array}{rrrrr}
	1&-1&&&\\
	&1&-1&&\\
	&&\ddots&\ddots&\\
	&&&1&-1\\
	-1&&&&1\\
\end{array} \right].
\end{equation}

%Since the sum of all columns (rows) of the matrix $\bs{A}$ yields a zero vector, the N columns (rows) are linearly dependent and $\bs{A}$ has one zero eigenvalue.
Since the vector $\bs{e}=[1,1,\dots,1]^T$ is an eigenvector of $\bs{A}$ to the zero eigenvalue, the determinant of the matrix vanishes. Consequently, the linear equation (\ref{eq:matrixeq}) has one free parameter $p\in[-1,1]$ (it behaves as en element of the vector $\bs{s}$, hence the constraint), which allows the solutions to appear, as shown below.

The system is easily solvable but leads to some constraints on what systems (e.g. having certain distributions of natural frequencies) can synchronize, which comes from the observation that $|s_i|\leq1$. The condition of solvability of the linear equation leads to the observation that the synchronized frequency is equal to the mean of the distribution of natural frequencies (the mean always exists in a finite population)
\begin{equation}
\Omega=\frac{1}{N}\sum^N_{i=1}\omega_i \quad.
\end{equation}

The synchronized solutions take the form of a set of phase differences between the neighbouring oscillators 
\begin{equation}
\label{eq:diffs}
\phi_{i}-\phi_{i-1} = \arcsin\left(p+\frac{2}{k}\sum^{N-i}_{j=1}\Delta_{j}\right) \quad,\, i=1,\, \ldots,\, N \quad,
\end{equation}
where $p\in[-1,1]$ is a parameter of the solution of the linear equation (\ref{eq:matrixeq}). For $i=N$ we assume that the sum equals $0$, i.e. $\phi_{N}-\phi_{N-1} = \arcsin p$.

Above, it is assumed that all inverse functions of sines $s_i=\sin \left( \phi_i-\phi_{i-1} \right)$ should be taken as $\phi_i-\phi_{i-1}=\arcsin s_i \in (-\pi/2,\pi/2]$ for the sake of stability of the solution (as discussed in Sec. 8 in more detail). For any of $N-1$ independent phase differences it is possible to take another inverse, $\pi-\arcsin s_i \in (\pi/2,3\pi/2]$, therefore in total one can obtain $2^{N-1}$ possibilities of how the solution can look like.

\section{The number of stable solutions}

\begin{figure}
\begin{minipage}[t]{\linewidth}
\begin{center}
  \includegraphics[width=0.8\hsize]{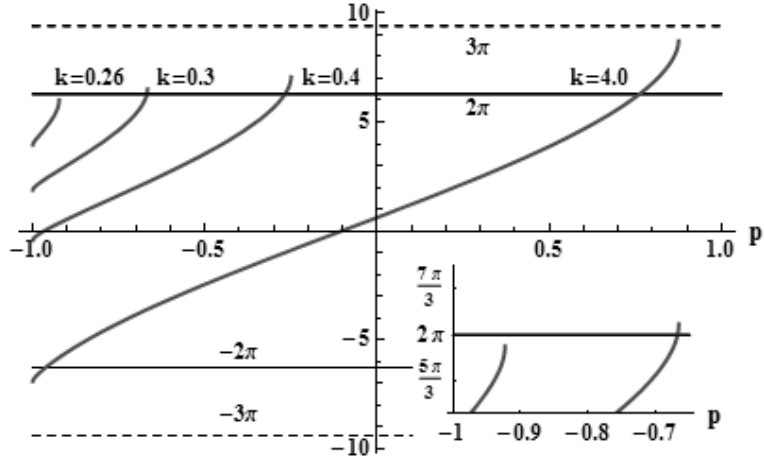}
  \caption{The illustration of the equation (\ref{eq:nawiniecia_0}) for $N=6$. Intersection of the curves (left-hand side of the equation) with the $0, \pm2\pi$ lines means that 0, 1, 2 or 3 solutions appear. The broken lines indicate limiting values obtained by the left-hand side.}
\end{center}
\end{minipage}
\end{figure}

For a given coupling $k$ and frequency distribution one finds stable solutions by solving for $p\in[-1,1]$ the equation obtained from summing up the $N$ phase differences (\ref{eq:diffs})
\begin{equation}
\sum^{N-1}_{i=0}\arcsin \left( p+\frac{2}{k}\sum^{i}_{j=1}\Delta_j \right)=2m\pi \quad,
\label{eq:nawiniecia_0}
\end{equation}
where $m=-\lfloor N/4 \rfloor, -\lfloor N/4 \rfloor +1, \ldots, -1, 0, 1, \ldots, \lfloor N/4 \rfloor$, which we call the "winding number" ($\lfloor\, .\, \rfloor$ denotes "`floor"', i.e. the largest integer number which is lower than or equal to the given number). Thus, there can exist more than one stable solution. Indeed, there can be at most $1+2\cdot \lfloor N/4 \rfloor$ synchronized solutions differing in winding numbers.

\begin{figure}
\begin{minipage}[htbp]{\linewidth}
\begin{center}
  \includegraphics[width=0.45\hsize]{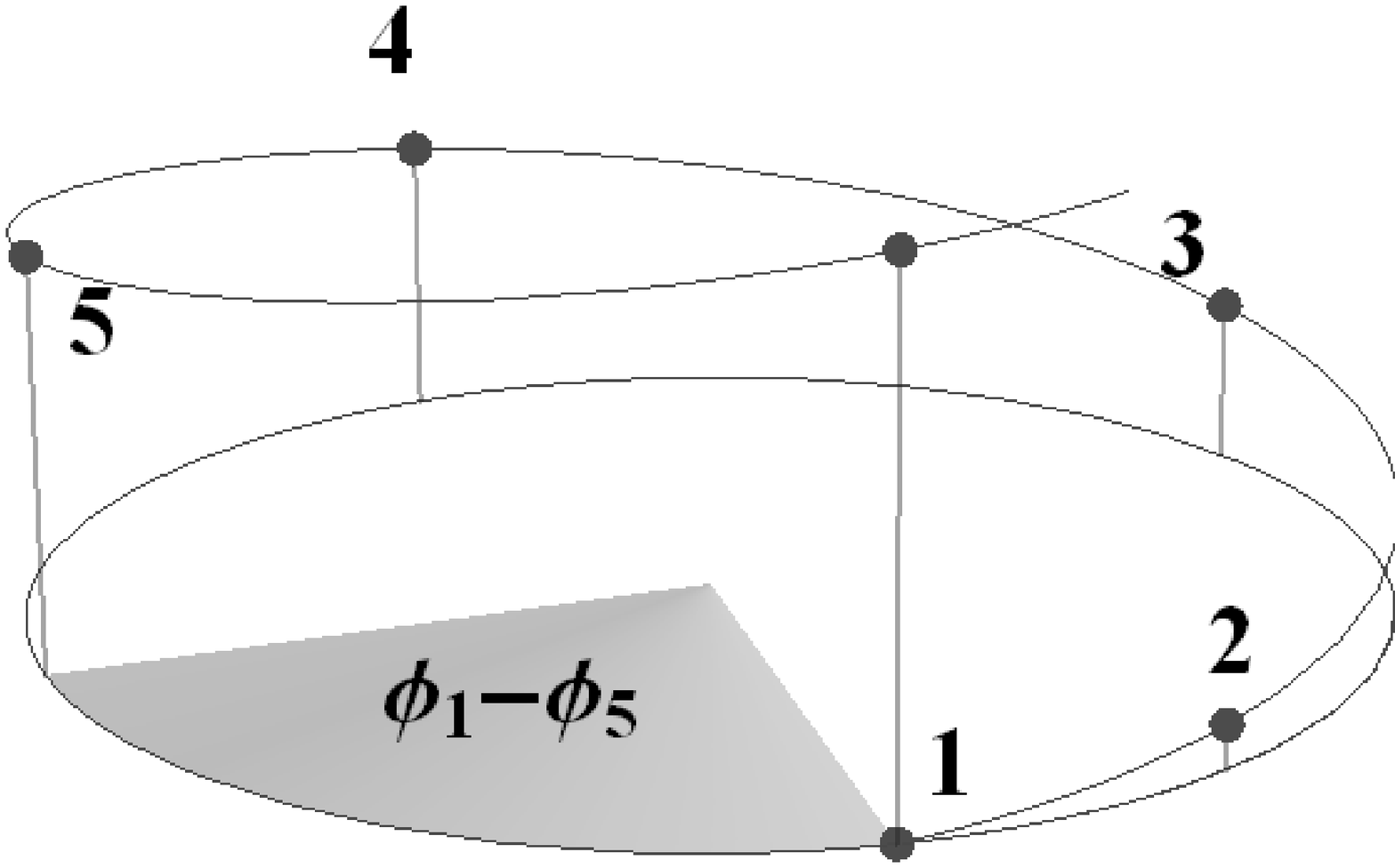}
  \includegraphics[width=0.45\hsize]{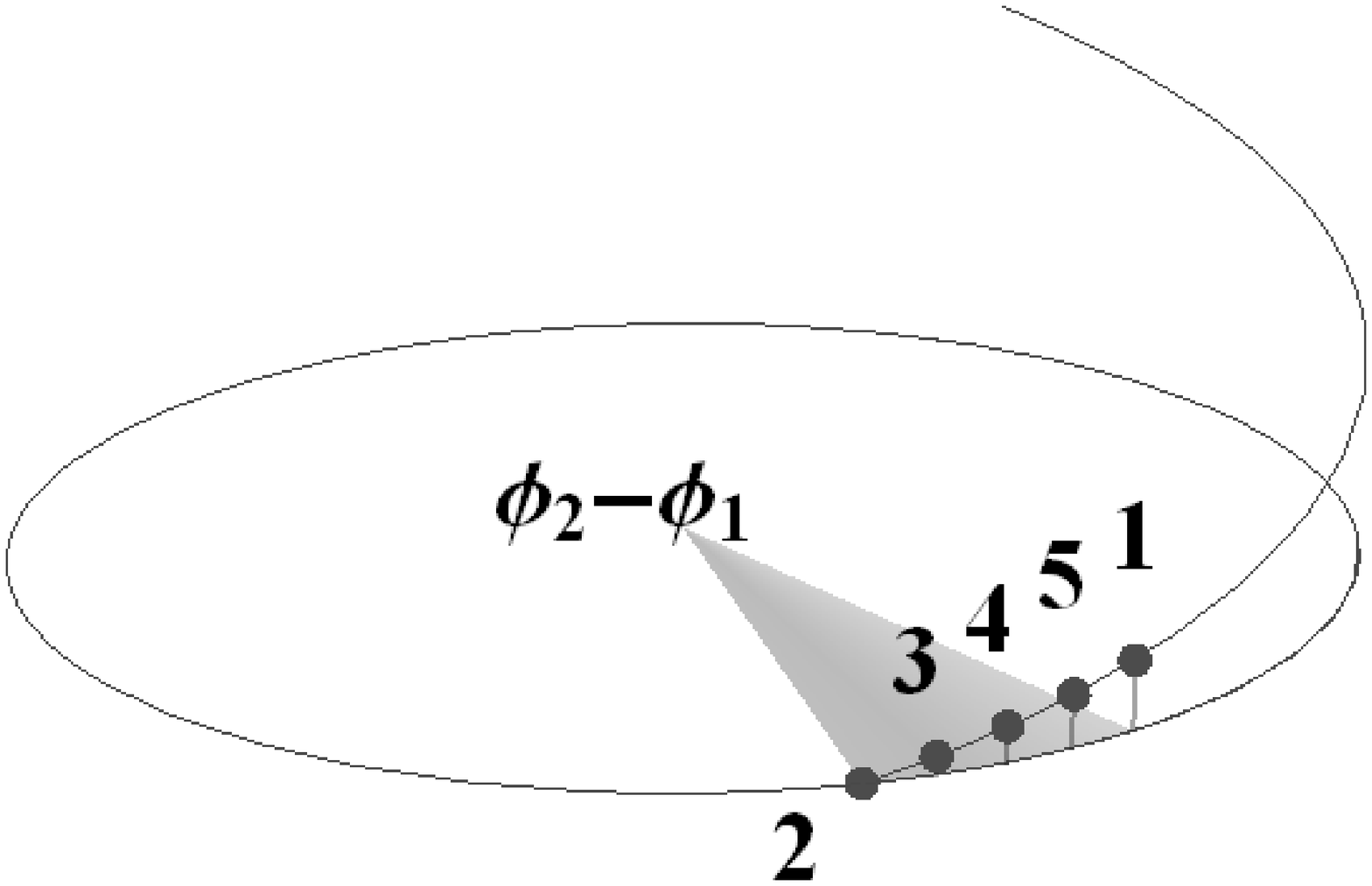}
  \caption{The idea of the winding number: the circle represents phase differences $\phi_{i}~-~\phi_{i-1}\in[0,2\pi)$. The digits are oscillators' indices. The right-hand solution has $m=1$, all the phase differences are positive and sum up to $2\pi$. The left-hand has $m~=~0$, one of the differences is negative and the sum is zero.}
\end{center}
\end{minipage}
\end{figure}

\section{Restrictions on synchronizing systems}
The form of interactions results in the fact that for synchronization to appear, all sums of deviations from the mean frequency must obey the inequality 
\begin{equation}
\label{eq:ineq}
\left|\frac{2}{k}\sum^{i}_{j=1}\Delta_j\right|<1 \quad,
\end{equation}
which binds the coupling to the natural frequency distribution. In the limit $N \rightarrow \infty$ independently of the frequency distribution, as shown by Strogatz and Mirollo \cite{Strogatz1}, the behaviour of the sums is given by
\begin{equation} 
\underset{1\leq i \leq N}{\max}|\sum^{i}_{j=1}\Delta_j| \sim N^{1/2} \quad.
\end{equation}
It comes from the fact that walking along the ring we can think of the subsequent random $\Delta_i$ as the steps of a one-dimensional random walk. This result was first obtained for Gaussian distribution but holds for any distribution of independently distributed natural frequencies.
%If the natural frequencies $+\omega$ or $-\omega$ are attributed to all oscillators with equal probability, the average distance of the walk (i.e. the sum of the deviations $\Delta_i$) scales as given above. 

It follows that if the critical coupling constant $k_c$ obeys the inequality (\ref{eq:ineq}), it is infinite in the limit $N \rightarrow \infty$. Consequently, infinite systems cannot synchronize, and so we restrict ourselves to the study of finite systems. The conclusion is compatible with the calculation of lower critical dimension by Daido \cite{Daido}.

The behaviour resembles that of the Ising model, which does not exhibit phase transition in one dimension, either (a brief discussion on the Ising model's solution on a ring as well as a commentary on the lack of phase transition in 1D can be found in \cite{Binney}). In the Ising model the phase transition occurs neither in infinite nor finite systems, while the one-dimensional Kuramoto model exhibits synchronization for a finite number of oscillators.

\section{Phase vs. frequency synchronization}
The order parameter for the original Kuramoto model,
\begin{equation}
	r(t) e^{i\psi(t)}=\frac{1}{N}\sum_{j=1}^{N} e^{i\theta_j(t)} \quad,
\end{equation}
indicates how well synchronized the oscillators are: if $r=0$, the system is totally incoherent, and if $r=1$, it is fully synchronized. It defines the critical coupling constant $k_c$ for which the continuous phase transition occurs: $r=0$ only for $k<k_c$, otherwise $r>0$.

In the local model the given definition does not serve its purpose. $r(k)$ provides the information on phase coherence which is a stronger condition than just frequency synchronization. Only the solution having a zero winding number can take values $r>0$ in the limit of large $k$. It results from the phase differences taking uniform values when $k$ increases (which can be seen already in the equation (\ref{eq:diffs})). 
%Only the solution having a zero winding number can reach $r\approx1$ for large $N$

\begin{figure}
\begin{center}
 	\includegraphics[width=0.6\hsize]{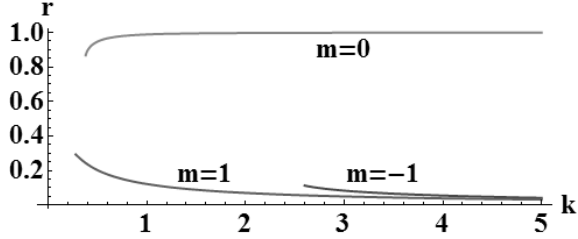}
  \caption{The order parameter $r(k)$. The synchronized solutions can be phase incoherent, hence $r\approx0$ for $m=\pm1$.}
\end{center}
\end{figure}

\section{The critical coupling}

It is interesting to find the critical coupling, where the phase transitions takes place for a given distribution of frequencies. Here, as we restrict ourselves to finite-size systems, $k_c$ just indicates the strength of interactions for which the first synchronized solution appears. To obtain $k_c$ for a given system (i.e. size $N$ and frequencies) one needs to calculate $\min \equiv \frac{2}{k}\min_{i=0,\ldots,N-1}\left( \sum_{j=1}^{i}{\Delta_j} \right)+1$, $\max \equiv \frac{2}{k}\max_{i=0,\ldots,N-1}\left( \sum_{j=1}^{i}{\Delta_j} \right)-1$ and next determine the values assumed by the left-hand side of the equation (\ref{eq:nawiniecia_0})
\begin{equation}
\left[\sum_{i=0}^{N-1}{\arcsin\left( \frac{2}{k}\sum_{j=1}^{i}{\Delta_j}-\min \right)}, \sum_{i=0}^{N-1}{\arcsin\left( \frac{2}{k}\sum_{j=1}^{i}{\Delta_j}-\max \right)} \right]
\end{equation}
The smallest value of $k$ for which one of the ends of the above interval equals $2m\pi$ is the $k_c$. This is tantamount to finding when (in terms of $k$) solutions appear, and choosing the first one.

\section{The stability of solutions}

We analyse the linear perturbation of the stationary solutions. Let $\{ \theta_i(t) \} ^N_{i=1}=\{\Omega t +\phi_i \} ^N_{i=1}$ be a synchronized solution of the system (\ref{eq:motion1}). We perturb it $\forall i:\theta_i(t) \rightarrow \theta_i(t)+u_i(t),$ where $ |u_i(t)|\ll 1$, and substitute it into the equations of motion (\ref{eq:motion1}). Having linearised the right-hand side with respect to $u_i$ we obtain the evolution equation of the perturbation
\begin{equation}
\label{eq:stab}
\dot{\bs{u}}\simeq - \frac{k}{2} \left[ \begin{array}{ccccc}
	c_1+c_2&-c_2&&&-c_1\\
	-c_2&c_2+c_3&-c_3&&\\
	&-c_3&c_3+c_4&-c_4&\\
	&\ddots&\ddots&\ddots&\\
	&&-c_{N-1}&c_{N-1}+c_N&-c_N\\
	-c_1&&&-c_N&c_N+c_1\\
\end{array} \right] \bs{u} ,
\end{equation}
where $\bs{u}=[u_1(t),u_2(t),\ldots,u_N(t)]^T, c_i=\cos(\phi_i-\phi_{i-1})$.

We consider the phase differences from the interval $\phi_i-\phi_{i-1}\in [-\pi,\pi]\,$ $(\mathrm{mod}\, 2\pi)$. Due to its cyclic structure, the  matrix in (\ref{eq:stab}) always has a zero eigenvalue, so that every solution is neutrally stable with respect to the homogeneous translations ("rigid rotations" of the system): $\forall i:\phi_i \rightarrow \phi_i + \tilde{\phi}$. All the other eigenvalues depend on the values of $c_i$.

We use the Gershgorin theorem (which can be found in \cite{Golub,Fortuna}), which states that for a matrix $\left[ a_{ij}\right]_{i,j=1,\ldots,N}$ all the eigenvalues lie in at least one of the circles (so-called Gershgorin discs) given by $\{ z:|z - a_{ii}| \leq R_i \}$, where $R_i~=~\sum_{j=1\\ j \neq i}^{N}|a_{ij}|$, $i=1,\,\ldots,\,N$. %[odnośnik do literatury?]

Using the theorem to localise the eigenvalues of the perturbation matrix we classify the solutions according to their  phase differences as follows:
\begin{description}
	\item[(i)] $\forall i=1,\ldots,N: |\phi_i-\phi_{i-1}|<\pi/2$ $\Rightarrow$ stable solutions (the eigenvalues $\lambda_i$ obey $\mathrm{Re}\lambda_i~\leq~0$)
	\item[(ii)] $\forall i=1,\ldots,N: |\phi_i-\phi_{i-1}|>\pi/2 \quad (\mathrm{mod}\, 2\pi)$ $\Rightarrow$ unstable solutions (the eigenvalues $\lambda_i$ obey $\mathrm{Re}\lambda_i~\geq~0$)
	\item[(iii)] $\exists i: |\phi_i-\phi_{i-1}|<\pi/2 \wedge \exists j: |\phi_j-\phi_{j-1}|>\pi/2 \Rightarrow$ we cannot determine the stability of the solution
\end{description}

The third condition describes the situation where at least one of Gershgorin discs intersects both the right and the left imaginary half-planes. The theorem does not tell us in which part of the disc the eigenvalues are localised, and whether any eigenvalue lies in it at all. In other words, there can exist stable solutions outside of $(i)$ and there can exist unstable solutions outside of $(ii)$.

\section{The size of basins of attraction}

The numerical evidence indicates that the stable solutions depend on initial conditions. The basin of attraction is a connected set containing the given solution. If it were otherwise, i.e. if the system reached a given stationary solution starting from a set of initial conditions which did not contain the solution, its trajectory would must have intersected a basin of attraction of another solution. As every point of the trajectory is identical to some initial condition, the system should have reached the other attractor, which is a contradiction.

We assume correspondence between the volume of the basins of attraction and the stability of solutions, that is the solutions having the maximal negative eigenvalue of the perturbation matrix closer to zero (i.e. they are "less stable") should have smaller basin of attraction than the "more stable" solutions whose all eigenvalues are strongly negative. It arises from the fact that for the system of first-order ODEs setting any initial condition non-identical to a given stationary solution is in fact perturbation of the solution. Since the negative eigenvalues indicate the speed of return to the solution, they also tell us which initial condition leads to which attractor. Of course, in the stability analysis we assume the perturbation to be small $|u_i(t)|\ll 1$, so the given argument should be treated as an assumption.

The simulations suggest that the value of winding number strongly correlates with both the eigenvalues and the volume of basins of attraction. The only analytical result states that the solutions with large $m$ are marginalised, because of the eigenvalues' behaviour
\begin{equation}
\lambda_i \sim -k \cos\left(\frac{2\pi}{N} m\right) \quad,
\end{equation}
where $m=-\lfloor N/4 \rfloor, -\lfloor N/4 \rfloor +1, \ldots, -1, 0, 1, \ldots, \lfloor N/4 \rfloor$. It is derived from the assumption that all the phase differences are equal, which is approximately the case for large $m$ and in the limit of large $k$ for any $m$.

\begin{figure}
\begin{minipage}[ht]{\linewidth}
\begin{center}
	\includegraphics[width=0.50\hsize]{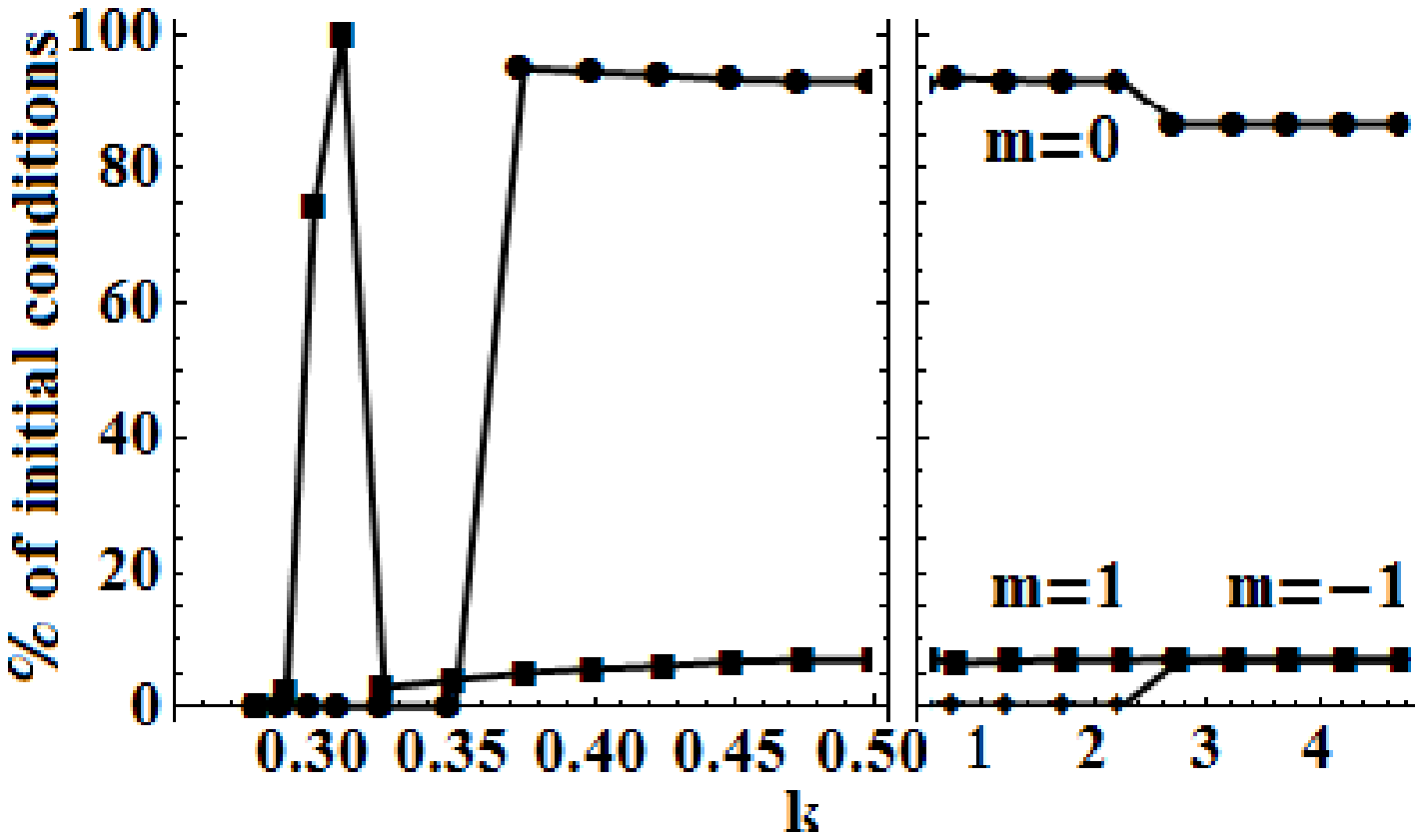}
	\includegraphics[width=0.49\hsize]{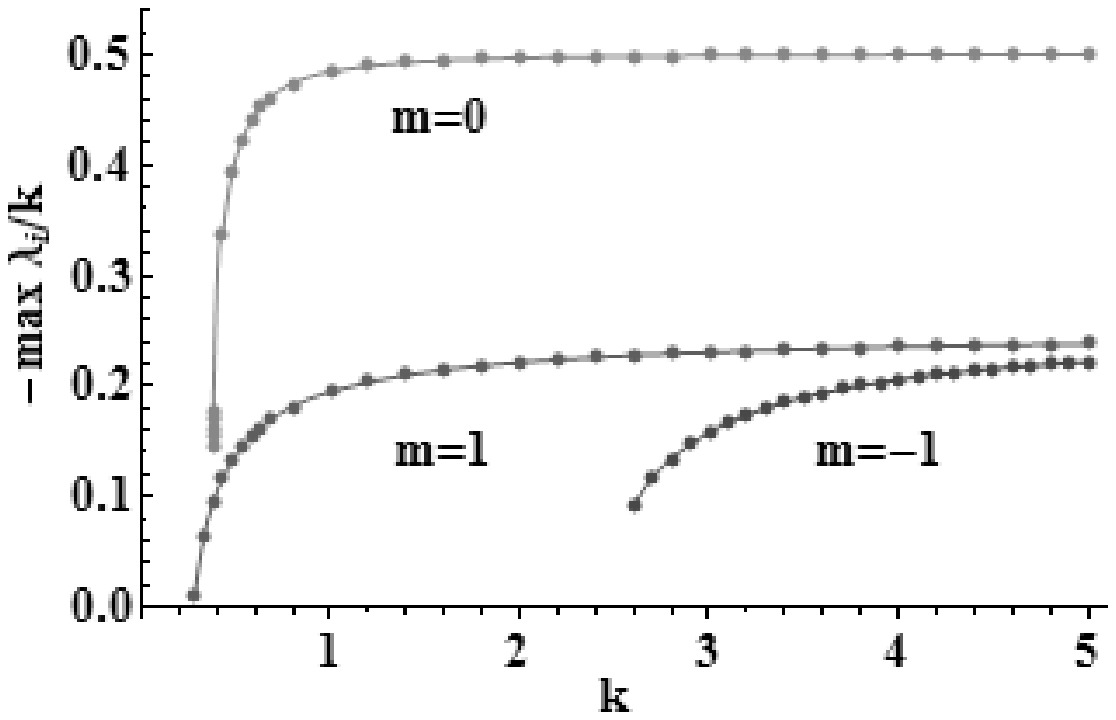}
  \caption{The basins of attraction and the eigenvalues of the perturbation matrix for solutions differing in winding number. Around $k=0.3$ the $m=1$ solution appears, then the one with $m=0$ takes over most of its basin of attraction, and at last, near $k=2.5$ the solution $m=-1$ appears building up its own basin with volume comparable to $m=1$.}
\end{center}
\end{minipage}
\end{figure}

\section{Conclusions}
As has been shown here, the one-dimensional model exhibits synchronization only for finite sizes of the system. The stationary synchronized solutions can be found explicitly, they depend on initial conditions, and their number is proportional to $N$. The size of the basins of attraction depends on the stability of the attracting solutions and, as we have shown, there is a connection between the winding number $m$ and the volume of the basin of attraction, according to which solutions with high $m$ are marginalised.

The order parameter for the Kuramoto model is not very useful here, as it indicates only phase coherence, while frequency synchronization may occur here without it. The critical coupling constant $k_c$ (which in finite systems considered here is understood as the smallest coupling for which any synchronized solution appears) can easily be computed. 

We hope that the analytically investigated behaviour of the one-dimensional model may give a clue, what to expect in the higher-dimensional local models of synchronizing oscillators.

%Synchronizacja może występować, choć jedynie w układach skończonych oraz dla pewnej klasy układów nieskończonych (które niestety eliminują losowość). [czy dopisać kawałek o nieskończonych?]

\end{document}